# Designing Autonomous Vehicles: Evaluating the Role of Human Emotions and Social Norms

Faisal Riaz[1], Muaz Niazi[2,*]

[1]Dept. Of Computing-Iqra University, Islamabad, Pakistan
[2]Dept. Of Computer Sciences-COMSATS, Islamabad, Pakistan
fazi_ajku@yahoo.com
*muaz.niazi@gmail.com

**Abstract**

Humans are going to delegate the rights of driving to the autonomous vehicles in near future. However, to fulfill this complicated task, there is a need for a mechanism, which enforces the autonomous vehicles to obey the road and social rules that have been practiced by well-behaved drivers. This task can be achieved by introducing social norms compliance mechanism in the autonomous vehicles. This research paper is proposing an artificial society of autonomous vehicles as an analogy of human social society. Each AV has been assigned a social personality having different social influence. Social norms have been introduced which help the AVs in making the decisions, influenced by emotions, regarding road collision avoidance. Furthermore, social norms compliance mechanism, by artificial social AVs, has been proposed using prospect based emotion i.e. fear, which is conceived from OCC model. Fuzzy logic has been employed to compute the emotions quantitatively. Then, using SimConnect approach, fuzzy values of fear has been provided to the Netlogo simulation environment to simulate artificial society of AVs. Extensive testing has been performed using the behavior space tool to find out the performance of the proposed approach in terms of the number of collisions. For comparison, the random-walk model based artificial society of AVs has been proposed as well. A comparative study with a random walk, prove that proposed approach provides a better option to tailor the autopilots of future AVS, Which will be more socially acceptable and trustworthy by their riders in terms of safe road travel.

## 1. Introduction

Autonomous road vehicles (ARVs) have been considered better than human driven vehicles in road safety and traffic management. According to the investigation of Riaz and Niazi [1], autonomous vehicles have been found helpful in decreasing road accidents as compared to the human-driven vehicles. Furthermore, the issue of road jams can be solved by replacing human drivers with fully connected autonomous cars, as noted by Litman [2]. In addition, Mersky and Samaras [3] have illustrated that AVs are very helpful in decreasing the road pollution and making the environment green. From these benefits, it might be possible that the law agencies delegate the driving task to AVs by issuing them the driving license. However, to fulfill this complicated task of driving, there is a need for a mechanism, which enforces the autonomous vehicles, which is a robot, to obey the road and social rules that have been practiced by well-behaved drivers.

The Role of ethics and social norms have been considered important in making robots social, well behaved and more compatible with humans. According to Malle [4], robots can serve as competent social agents by integrating moral norms in their basic architecture. A. Rakotonirainy et al. [5] have proven that social norms can be utilized to design human compatible social AVs robots. According to Kummer et al. [6], social norms can be used in tailoring crash free AVs robot by operating on roads wisely. From the above discussion, it is implied that social norms with some norms compliance mechanism can be used to tailor the next generation of more trustworthy social AVs.

Emotions can be used as norms compliance mechanism as it is already proven that emotions help in sustaining the social norms in human society. According to Elster [7], self-attribution emotions like shame helps the human to avoid the violation due to fear of losing their social status. According to N. Criado [8], prospect based emotions like fear enforce the human to follow the social norms in order to avoid the punishment from the law enforcement agencies. Inspired by the role of emotions in social norm compliance in the human society, researchers have used emotions to enforce the artificial agents' norm compliance. According to staller et al. [9]emotions act as an important factor in the sustainability of social norms. Hence, it is implied that we can use emotions as the norms compliance mechanism to design social norms enabled AVs. Gerdes and Thornton [10] have suggested the mathematical model of social norms for designing the control algorithms of AVs but still their works lack the simulation or proof of concept of proposed mathematical models.

**Problem statement**- However, to the best of our knowledge the existing literature has not proposed such procedures that allow AVs to configure their autopilots to make collision avoidance decisions about norm compliance using emotional motivations as human drivers would do. For example, Amitai and Oren[11], just suggested the use of social norms in AVs in the theoretical aspect without discussing any working mathematical model and its implementation aspects. A. Rakotonirainy et al. [5] have been proposed a novel concept of measuring the emotional state of a driver using the HUD-UP technology and transmitting the social norms from driver to driver to modify the behavior behind the steering of AV. However, the concept of social norms has not been integrated into the autopilot of the AV, which helps them to make the collision avoidance decisions by their own. The major challenge for AVs is that how it will take decisions at the time of the crashes and this issue has been addressed by the Kumfer and Burgess in [6]. The authors have used social norms as a decision mechanism to choose a less harming crash among possible collision options. However, this paper does not provide any collision avoidance strategy using some social norms compliance mechanism, which avoids the collision situations.

**Contribution -** The existing research work is proposing a set of contributions in building the norm compliance collision free artificial society of Autonomous vehicles inspired by human society social norms and related emotions. Our aim is to provide humans with reliable AVs to which they can delegate driving tasks that are regulated by legal and social norms. The main contributions of the paper are given as.

- *Viewpoint of incorporating the social norms and emotions in Autonomous vehicles and conceiving them as artificial social entities*
- *Modeling of social norms inspired artificial society of Autonomous Vehicles*
- *Modeling of prospected based emotions to make AVs emotions enabled*
- *Simulation of social norms compliance artificial society of AVs using the Net logo*
- *Detailed experiment design*
- *Rigor analysis ,in terms of number of collisions, of the proposed approach in the comparison with random walk travelling strategy*

This article is organized as follows: Section 2 illustrates the literature review; Section 3 describes the method; Section 4 provides the description of the proposed model; section 5 illustrates the experiments; section 6 elaborates the results and discussion and section 7 contains conclusion.

## 2.-Literature Review

In this section, detailed literature review related to the proposed scheme has been performed. The literature review has been divided into three main categories. The first category addresses the literature supporting the role of ethics in robots using theoretical debate. The second category discusses the literature that supports the role of using ethics or norms in the design of AVs but only theoretically. Then the third category discusses the state of the art literature, which has used the social norms in autonomous vehicles using a simulation approach.

According to Voort et al. [12] computers are getting autonomous day by day and are capable of making decisions of their own. Intelligent computer systems can get information from human, analyze it, take decisions and store that information or provide it to third parties. There is a need to check the moral values of computer decisions. Authors have suggested that there is a need to add ethics in technology, which is still lacking behind. Malle1 [4] summarized that from 1995 to 2015 very little efforts have been made on the implementation of ethics in robots. Past studies provide a thought that whether a robot could be a moral agent or not. In addition, researchers found that robot could be treated as a living thing that can take actions on its own decisions, and it can decide what is right and what is wrong with humans.

According to Amitai and Oren [13], the latest smart machines like AVs are getting smarter due to the incorporation of artificial intelligence (AI) algorithms. Furthermore, these AVs are becoming more and more autonomous in the sense that they are now taking decisions on their own using these AI algorithms. The authors suggest that as these AVs basic purpose is to serve humans, and then there is a need to equip them with ethical and social rules so that the autonomous devices like AVs can take decisions of their own that could not harm the passengers and other road commuters [7]. According to Gerdes and Thornton [10], it is the responsibility of researchers and programmers to devise ethics enabled control algorithms for AVs that make them more acceptable to human society. The authors argue that the incorporation of ethics of the society in which AVs are operating will help the court of law to decide the responsibility level of AV in the case of an accident. In this regard, they have proposed a mathematical model of ethical frameworks to incorporate them in the control algorithms of Autonomous vehicles. The proposed model can read the error rate for the actual and desired path of the car based on different constraints. However, the authors have not mentioned any case study that implements any of the proposed mathematical model using simulation or real field tests.

Social and autonomous robots are the motivation to build social cars so that road accidents can be eliminated. Vehicle-to-vehicle (V2V) is a sub-part of intelligent transportation system (ITS) equipped with sensing technologies and wireless communication system is helpful in road accident prevention. A. Rakotonirainy et al. [14] have been proposed a novel concept that HUDs , Human-Computer-interaction, HCI and communicating social information between cars can provide social awareness and he named it as 'social car'. This social car can sense the driving behavior of driver by capturing the facial expression, gesture and eye contacts of the driver. Further, the author has argued that self-efficacy and social norms can change the driver's behavior. Social norms can be transmitted in V2V using social networks and most of the time in the form of non-verbal communication. Hence, the combination of driver and car (machine) become the cyborg so the driver of one can treat the other driver as a machine. He also added that the "social pressure is particularly suitable to influence human driving behaviors for the better and that this aspect is still relevant in the age of looming autonomous cars".

A complete autonomous vehicle (AV) was introduced first time in response to Defense Advanced Research Projects Agency Grand Challenges. The major challenge for AVs is that how it will take decisions at the time of the crashes. This is the key point where ethics and social norms are required for AV's development. To address this requirement the Kumfer and Burgess [15]evaluated three ethical theories, i.e. utilitarianism, respect for persons, and virtue ethics, which help AVs to make least harming collision decision when the collision become unavoidable. They performed the experiments using MATLAB their results revealed that the utilitarian system produced the lowest number of death while on the other hand, the virtue ethics system resulted in a supreme number of losses. However, the virtue ethics if fully integrated with good AI techniques can be the best ethical solution. It is suggested that these ethical theories can be implemented in AVs in different scenarios and complex environments.

**3-Method**

This section presents the method that has been used to propose the social norms and emotions inspired artificial society of AVs. Figure 1 is the pictorial representation of our proposed method. To introduce the emotions, a suitable appraisal model was required. According to [25], OCC model is a best emotion appraisal model. Hence, the OCC model has studied thoroughly and Prospect-based emotions have been selected. Further, emotion Fear has been selected to devise the mechanism, which enforces the agents to obey the social norms in different collision leading road scenarios. Afterward Fuzzy logic has been employed to compute the quantitative values of different intensities of fear. Then social norms and emotions based rules have been designed, which define the code of conduct for the artificial society of AVs. In addition, artificial social actors along with different characteristics have been defined. Now to test the behavior of non-social norms and social norms based artificial society, standard agent-based modeling tool NetLogo has been used. Using, Sim-connector approach the numeric values of Fear emotion have been provided in the simulation of artificial society. Then extensive experiments have been performed, which helps to perform the comparison between non-social norms and social norms compliance artificial society of AVs. Furthermore, the method of using prospect-based emotion for generating emotions in agents has been compared with existing work in literature.

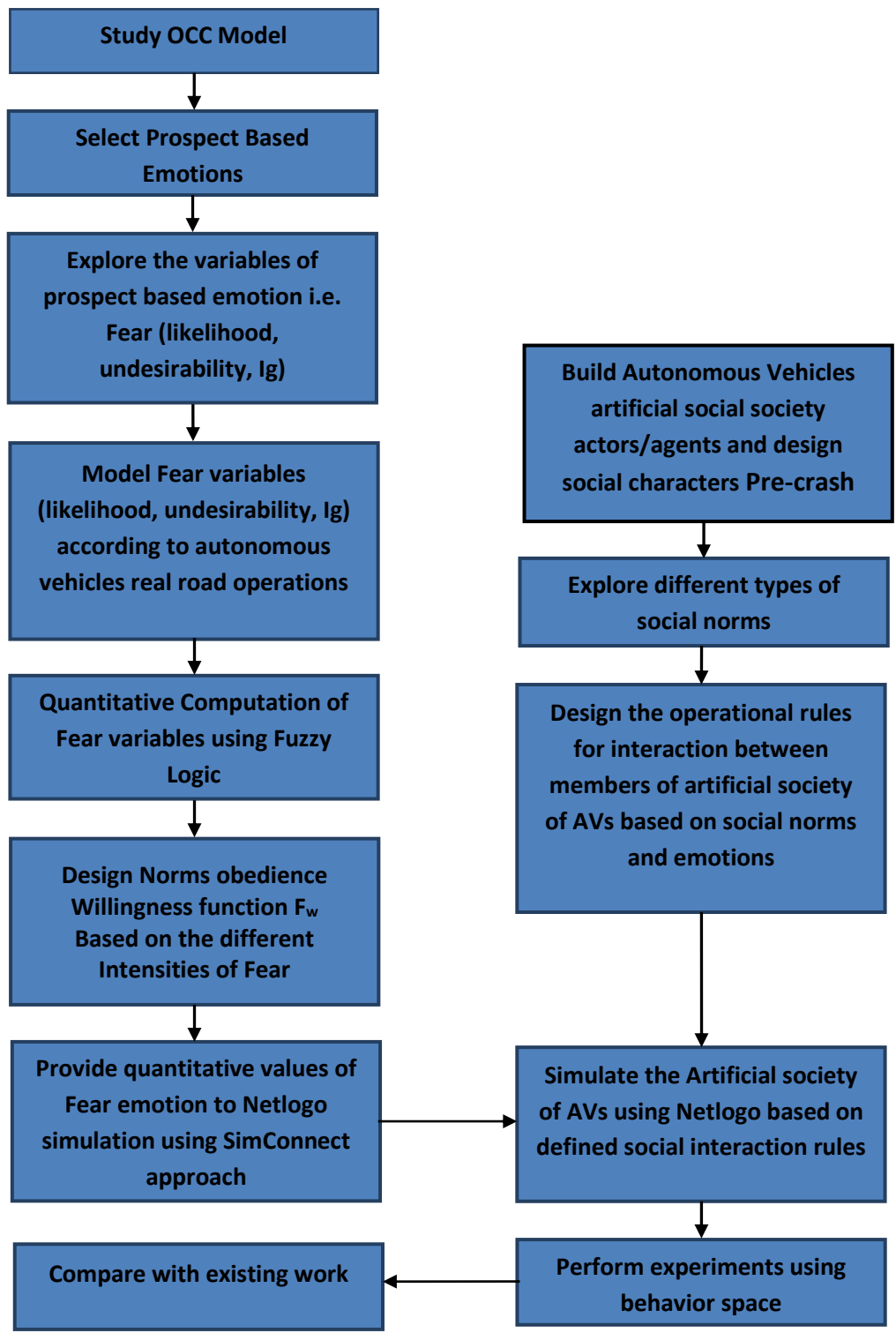

Fig 1. Proposed Method

# 4-Description of the Novel Solution (A view point)

In this section, the detailed description of the proposed viewpoint and model has been presented.

## 4.1 Humanizing the AVs: A Possible Design Decision?

In order to include the social norms and emotions in the AVs, their fundamental design needs to be changed because it is important since it can lead to more human-like behavior on the future roads. It is important to note that emotional action is a social action that helps to regulate and adapt other actors' emotions and emotional expressions according to valid norms and rules [16]. The emulation of emotions and social norms in the design of AVs can help in building a possibly more comfortable, trustworthy and collision free AVs. However, the exact mode of implementation is still debatable. The supposition guiding this method would be that the rules and norms guiding human-human interaction/communication might also be pertinent in AV design. Thus, the design principle would have a goal to introduce anthropomorphism capabilities in AVs (Fig. 2).

## 4.2 How emotions enforce social Norms in Autonomous Vehicle: A scenario

To analyze the above discussion further, let us examine a conjectural interaction situation of two autonomous vehicles as shown in figure 3. Suppose "*A*" is a heavy autonomous truck followed by a smaller-sized autonomous vehicle "*B*" considerably less in weight but occupying the same lane. Here, one way of representing "*A*" could be to consider it as a strong influencing person having a strong social status, whereas "*B*" is a less influencing person having a weak social character. For safe driving, both actors have to follow social norms and rules. In different social societies of the world, weaker feels the emotion of fear from stronger, whereas stronger can have multiple emotions for the weaker like sympathy and pride, etc. Actor "*B"* should maintain a safe distance from actor "*A"* or actor "*A"* should practice sympathetic emotion for "*B"* to avoid the collision. Let us suppose," *A"* decreases its speed without taking care of in-between distance from "*B"*, which might lead to the collision. Suddenly another actor "*C"* (having strong social status) appears on the road. "*A*" starts feeling the fear that if it collides with "*B",* then "*C"* will have the evidence of its cruelty. According to a social norm, "***You will be punished for your act of crime*** "generates an emotion of fear of losing its status in social society and get punished from the law enforcement institutes. Consequently, *A* maintains a safe distance from *B* and avoids the collision.

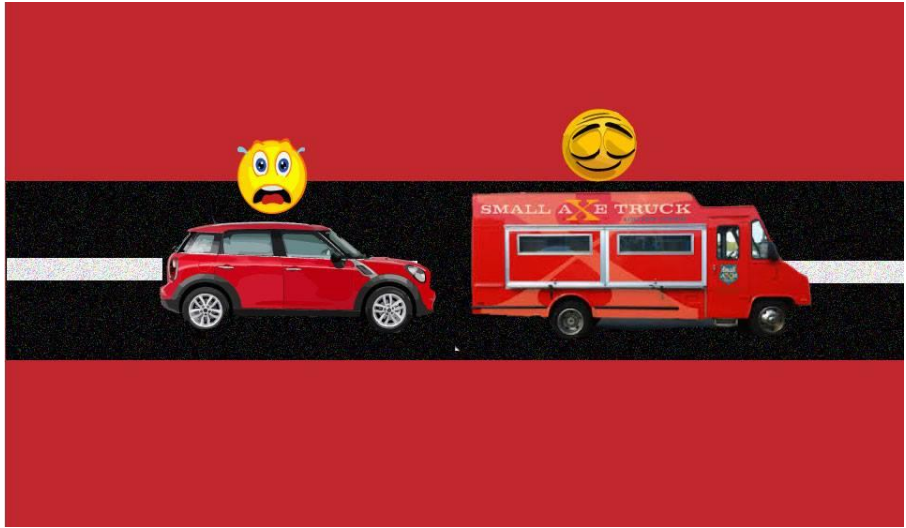

Fig 2. Autonomous vehicles exhibiting different emotions based on their size and situation

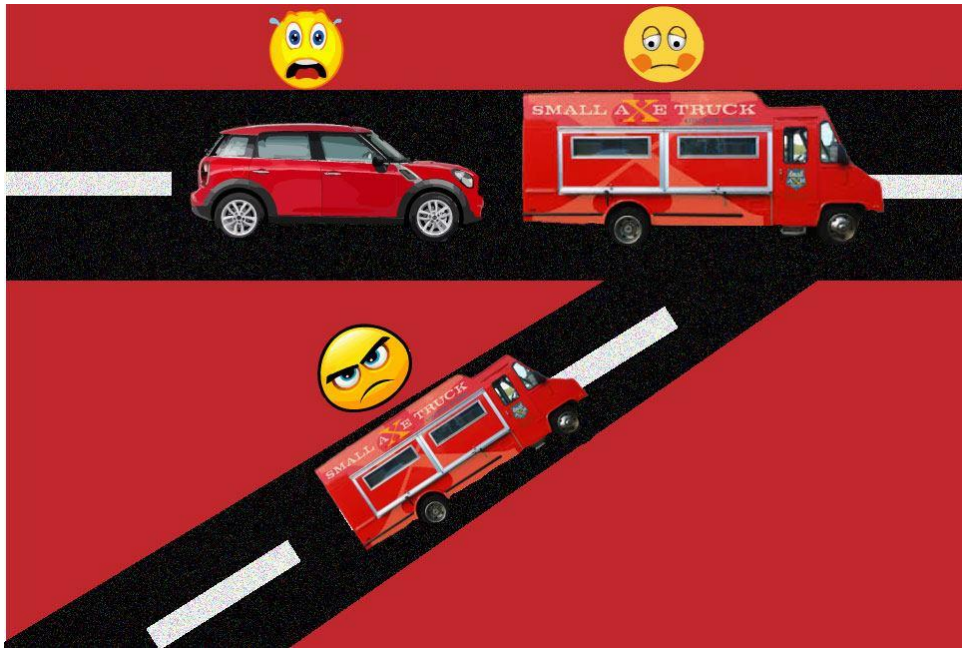

Fig 3. Collision avoidance scenario using emotions enforced social norms

What has happened here? The primary event of fretting the weaker autonomous vehicle can be defined by some appraisal theory. This is the social norm that "Any actor of any status will be punished for performing evil deed". Therefore, actor "*A"* avoids or feels hesitation in the execution of collision scenarios in the presence of witnesses. This social norm generates the fear of being punished. It means fear forces the actor to follow the norm.

**4.3- Artificial Social Society of AVs**

To further elaborate the idea presented in the previous section, we have given a concept of an artificial society of AVs, consists of different actors, and having each one different characteristic, which depicts their different social personalities. These different actors having different personality characteristics along their short abbreviated nicknames are presented in table 1. These characteristics have been assigned to these AVs from the real life behavior of human drivers driving these types of non- autonomous vehicles. It has been observed that drivers of heavy vehicles act dominantly, do not let the lighter vehicles to overtake or treat them harshly, in the real road traffic.

TABLE 1
DIFFERENT ACTORS OF ARTIFICIAL SOCIETY OF AVS

| Actor | Symbol of Actor | Personality Type | Weight |
|---|---|---|---|
| AV_Truck | T | Very Dominating | >5000 |
| AV_Bus | B | Very Dominating | 4500<∧<50000 |
| AV_Toyota_Small truck | TY | Dominating | 3000 ∧<4000 |
| AV_Carry | C | Weak Dominating | 2000 ∧ <2800 |
| AV_Car 3000cc | CB | Weak Dominating | 2000 ∧ <2500 |
| AV_Car 2000cc | CS | Weak Dominating | 1500 ∧ <1700 |
| AV_Rickshaw | R | Weak Dominating | 1200 ∧ <1400 |
| AV_Ambulance | A | Weak Dominating | 1200^<1400 |
| AV_Motorbike | M | Very weak Dominating | 800 ∧ <1100 |
| AV_Cycle | CL | Very weak Dominating | 400 |

*4.3.1 -Operational rules for artificial society of AVs*

Humans live socially following some rules, which help them to live peacefully, avoiding any possible conflicts, if follow them properly. In the analogy of these social rules, we have proposed some rules as well for the proposed artificial society of AVs. These rules have been designed in the light of different possible road scenarios, though all cannot be mentioned here, and further, the social norms have been presented as well along with the best-suited emotions. Furthermore, to check weather following social norm, according to the given condition, does not lead to the violation of the road traffic rules, we have applied a check to assure that the social norm will be followed only when the social norm and road norm both are in the compliance of each other. The last column of table 2 presents the action that the actor has to be taken to avoid the road collision.

We have used both formal and informal social norms, in our proposed model, and these have been rewritten in the context of AVs collision avoidance strategies. For example:

- You ought to maintain a safe distance from stronger vehicles to avoid the collision.
- You ought to be kind to the weak vehicles to avoid the collisions
- Keep on your Lane
- Help in executing safe overtaking maneuver

TABLE 2
SOCIAL NORMS AND EMOTIONS BASED ROAD INTERACTION RULES

| Road Scenarios | Social norm (1=Self-norm 2=Religious norm 3=Enforced Norms) | Emotion | Road norm (P) ||(p &q) | Action |
|---|---|---|---|---|
| T is leading the CB and CB wants to overtake the T | Help the weaker | Sympathy, Guilt, Shame | (Should assist overtaking vehicle to get past swiftly and securely) & (Keep your lane). | P & q -> social norm || p & q !-> social norm |
| CB is leading the T and T wants to overtake the CB | Maintain distance from stronger one | Fear | (Must not drive in a bus lane) & (Should assist overtaking vehicle to get past swiftly and securely) | P & q -> social norm || p & q !-> social norm |
| CB–CB following scenario and the following CB wants to overtake the leading CB. | Give the right or Tit for Tat | Guilt, shame, Sympathy | (Must not drive in a bus lane) & (Should assist overtaking vehicle to get past swiftly and securely) | P & q->social norm || p & q !-> social norm |
| T is following the B and T wants to overtake B | Give the right or Tit for Tat | Fear ,guilt ,shame | (Increase your speed to keep a safe distance from other vehicles) &(Should | P&q -> social norm || p&q !-> social norm |

| | | | | |
|---|---|---|---|---|
| | | | assist overtaking vehicle to get past swiftly and securely) | |
| If CB following T in bright conditions | Keep the distance from stronger one | Fear | Keep the safe distance and use the two-second rule. | P -> social norm \|\| p!->social norm |
| If CB following T in rainy conditions | Keep the distance from stronger one | Fear | Keep the safe distance and use the two-second rule. | P -> social norm \|\| p !-> social norm |
| If CB is involved in tailgating with T | Keep the distance from stronger one | Fear | (Increase your speed to keep a safe distance from other vehicles ) | P -> social norm \|\| p!->social norm |
| If T is involved in tailgating with CB | Keep the distance from stronger one, Abide the rule | Fear | (Increase your speed to keep a safe distance from other vehicles) & (must not drive in a bus lane) | P & q -> social norm \|\| p & q !-> social norm |

## 4.3.2-Generate Prospect-based Emotions using OCC Model

A very popular model of emotions was developed by Ortony, Clore and Collins also known as OCC model [17]. The reason for choosing the OCC model is they are presenting 22 basic emotions along with the concept of computing the emotions as well. In OCC model, the authors answer the question that what concludes the strength of the emotions. They commence a number of variables in enjoining to answer this problem. To address this problem, they introduce three types of variables as shown in Table 3. The OCC model is shown in the figure below. The variables given in Table 3 help to compute the strength of the emotions. However, these values are still computed in qualitative manners. However, we will compute their quantitative values using fuzzy logic first and then use them to evaluate proposed schemes given in coming sections.

### 4.4 Overall functionality of proposed approach

The overall functionality of the proposed approach is presented in figure 4 and its description is given as under.

1. In EventPart1, CB is following T and it requests the T to give safe passage for performing an overtaking maneuver. Belief will depict the situation awareness of CB and T. The current values of Belief are passed to the Prospect Based Emotion Generation module.

2. Prospect Based Emotion Generation module computes the emotion fear based on equation 1 and 2 provided by OCC model [25].

$$Fear - Potential(Av_i, e_i, t_i) = f_{f[\ |Desire\ (Av_i,e_i,t_i)|,\ Liklihood\ (Av_i,e_i,t_i),\ I_g(Av_i,e_i,t_i)]} ..(1)$$
$$Fear - Intensity\ (Av_i, e_i, t_i) = Fear - Potential\ (Av_i, e_i, t_i) - Fearthreshold\ (Av_i, e_i, t_i) .. (2)$$

3. The computed Intensity of fear will update Belief of the agent.
4. Based on Intensity of fear $F_w$ is computed. The Fw function is computed using equation 3. Equation 3 shows that the willing function of T, which allows the overtaking, depends on the intensity of fear.

$$F_{w\ (obey-norm)} = Fear - Intensity\ (Av_i, e_i, t_i) .. (3)$$

In the first iteration, the value of the emotion will be zero. In the Willing Function part, it will be checked whether the value of Willing Function of T is higher than λ or it is less than λ. Here two scenarios exist; first, if the $F_w$ of T is less than λ, and second if the $F_w$ is greater than λ.

5. If the value of $F_w$ is smaller than the egoist value of agent then it disobeys the Norm.

6. In the case of disobeying the Norm, T will be entered in the pre-crash scenario. For the pre-crash scenario we have considered the variables defined by [18]

7. The event of pre-crash scenario will contribute in the shape of the High likelihood of an accident and it will increase the intensity of fear. Again, the belief of agent will be updated and $F_w$ will be computed. If the $F_w$ is still smaller than $\lambda$ then the emotion generation center will be consulted again to depict a highly dangerous situation. If the value of $F_w$ is greater than $\lambda$ then the egoist agent will change its mind and turn into the emotional agent and the emotions act as a norm compliance mechanism. In next step agent will check the road norm that accepting the preceding AV (CB) request is not against the road norm, then it will load possible solutions and execute the maneuver, which will ultimately help in CB in performing an overtaking maneuver by avoiding rear end collision.

**5-Experiments**

This section describes the two types of experiments: Quantitative Computation of Prospect-based Emotion using Fuzzy Logic and validation of EEC_Agent .

### 5.1 Experiment 1

Since human emotions are fuzzy and complex in nature, using fuzzy sets for modeling the human emotions can be a suitable choice [19]. For modeling the emotions, fuzzy logic has been extensively utilized [20]. A computational model of emotions has been proposed that can be included in any cognitive agent or program. In [21], the Takagi-Sugeno (TS) fuzzy model has been utilized to develop an online learning system of emotions. The purpose of designing this system was to investigate that how multi-model actions can be generated and understand by cognitive robots. In [22], a novel method is proposed which helps to model the emotions using different types of physiological data. For this purpose two fuzzy logic models are employed: one model is used for converting the signals into valence and arousal and the second model is used for converting this valence and arousal to five emotional states related to the computer games e.g. boredom, excitement, challenge, frustration, and fun. In our case, the OCC model provides a computation traceability algorithm shown below for computing the intensity of fear in [23].

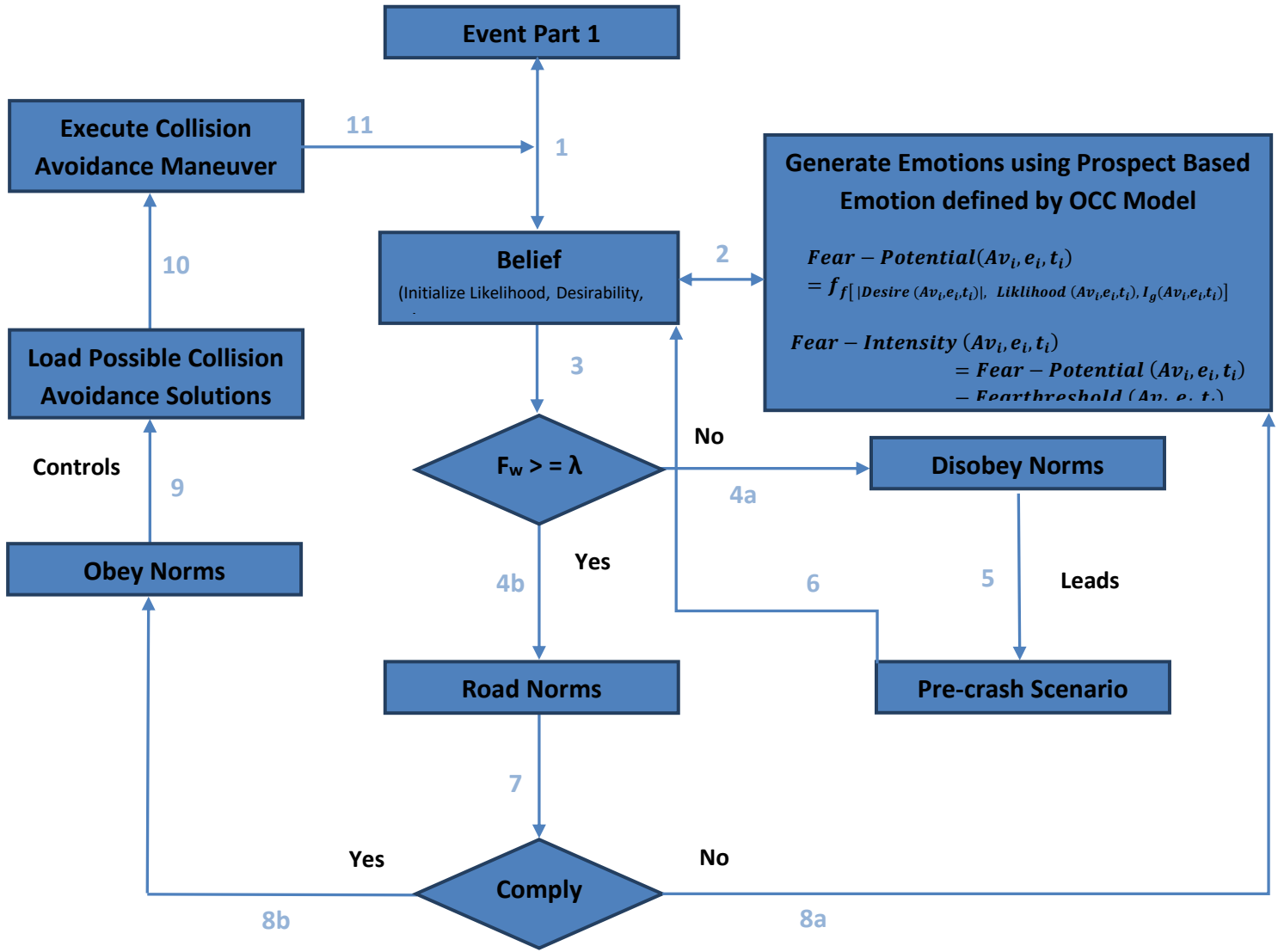

Fig 4. Overall functionality of proposed approach

However, it is useless if the numeric values of linguistic variables like likelihood, desire and Ig variables are not known.

If Prospect (v, e, t) and Undesirable (v, e, t) < 0

Then set Fear-Potential (v, e, t) =  ff [|Desire (v, e, t) |, Likelihood (v, e, t), Ig  (v, e, t)]

If Fear-Potential (v, e, t) > Fear-Threshold (v, t)

Then set Fear-Intensity (v, e, t) = Fear-Potential (v, e, t) - Fear-Threshold (v, t)

Else set Fear-Intensity (v, e, t) =0

*5.1.1 Implementation details of fuzzy logic to compute the numeric values of Fear Emotion*
In order to compute the Fear-Potential as given in part 1 of the computation traceability algorithm, we have to calculate the values of Desirability, Likelihood, and Intensity of a global variable. In the context of the state of the art given above, we used fuzzy logic to compute the numeric values of Desirability, Likelihood, and Intensity of global variable.

  a) *Likelihood:*

For the variable of Likelihood the five linguistic tokens VLLH, LLH, MLH, HLH and VHLH were defined which represent Very low likelihood, Low likelihood, Medium likelihood, High likelihood and Very High likelihood respectively shown in table 3.

TABLE 3
LIKELIHOOD LINGUISTIC TOKENS AND THEIR DESCRIPTION

| Linguistic Tokens | Description |
|---|---|
| VHLH | Very High Likelihood |
| HLH | High Likelihood |
| MLH | Medium Likelihood |
| LLH | Low Likelihood |
| VLLH | Very Low Likelihood |

The variable of likelihood is affected by Distance and Speed variables. Twenty-five rules were defined to obtain the value of the variable likelihood; these rules are given in the following table.

TABLE 4
LIKELIHOOD FUZZY INFERENCE RULES

| If Distance is | And Speed is | Then Likelihood is |
|---|---|---|
| VHD | VHS | MLH |
| VHD | HS | LLH |
| VHD | MS | VLLH |
| VHD | LS | VLLH |
| VHD | VLS | VLLH |
| HD | VHS | HLH |
| HD | HS | MLH |
| HD | MS | VLLH |
| HD | LS | VLLH |
| HD | VLS | VLLH |
| MD | VHS | VHLH |
| MD | HS | VHLH |
| MD | MS | MLH |
| MD | LS | LLH |
| MD | VLS | VLLH |
| LD | VHS | VHLH |
| LD | HS | VHLH |
| LD | MS | HLH |
| LD | LS | MLH |
| LD | VLS | VLLH |
| V LD | VHS | VHLH |
| V LD | HS | VHLH |
| V LD | MS | VHLH |
| V LD | LS | HLH |
| V LD | VLS | MLH |

The remaining details are given in appendix A.

## 5.2 -Experiment 2

The purpose of the second main experiment is to simulate the concept of an artificial society of AVs, which consists of different actors having different characteristics. Another reason of simulation is to study the behavior of these actors according to the defined social rules during autonomous driving. For this purpose, Netlogo 5.3 has been utilized which is a standard agent-based simulation environment. The NetLogo environment consists of patches, links, and turtles [24]. The algorithms used in this experiment have already been given in section 3.1.3. Figure 5 presents the experimental environment along with input and output parameters. The left side of the simulation world contains input sliders for providing fuzzy logic based numeric values of prospect based emotion variables (Undesirability, Likelihood, Ig). It is important to recall here that these numeric values of prospect-based emotions were computed through experiments a, b and c using fuzzy logic and then provided to the agent based simulation by following proposed SimConnector approach.

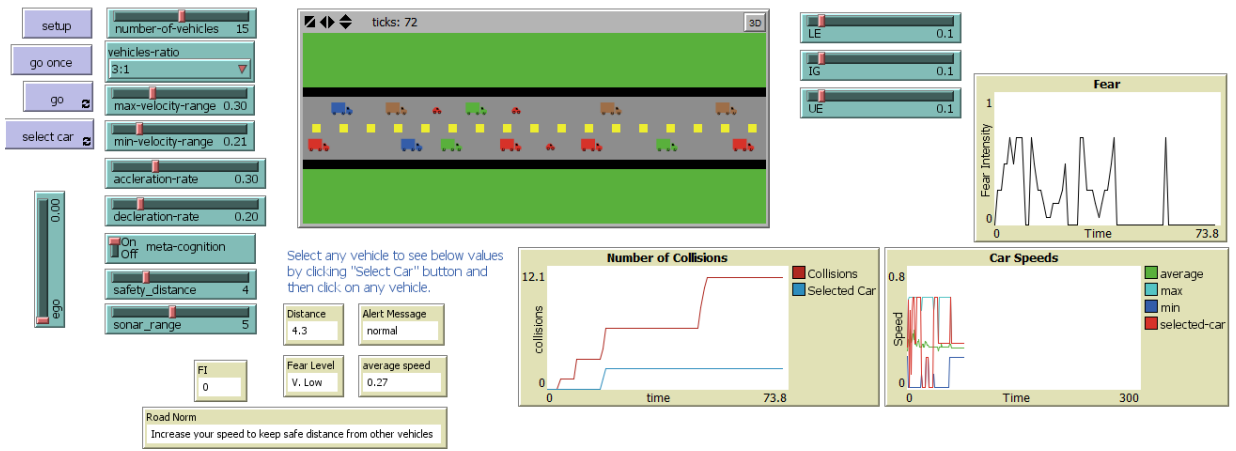

Fig. 5 Main Simulation Screen of Social Norms and Emotions inspired artificial society of AVs

*5.2.1- Simulation Parameters Description*

The simulated world consists of different types of input and output parameters. To provide the inputs, different sliders have been used, whereas to get the outputs, monitors and plots are used. The description of each input and output object along with defined range is presented in table 5.

TABLE 5
SIMULATION PARAMETERS AND THEIR DESCRIPTION

| Simulation General Parameters | Range | Description |
| --- | --- | --- |
| Number of Autonomous Vehicles Agents | [1-30] | This slider helps in defining the maximum members of the artificial society of AVs. |
| Vehicles Ratio | [2:1, 3:1, 4:1] | It defines the ratio of AV Trucks and cars within a total number of vehicles set by the Number of Autonomous Vehicles Agents slider. |
| Maximum Velocity | [0 -1; with increment of 0.01] | This slider helps in defining the maximum velocity that can be achieved by all actors of artificial society |
| Minimum Velocity | [0 -1; with increment of 0.01] | This slider helps in defining the lower boundary of velocity achieved by all actors of artificial society |
| Acceleration-rate | [0 -1; with increment of 0.05] | This slider helps in defining the maximum acceleration rate that can be used by all actors of artificial society |
| Declaration-rate | [0 -1; with increment of 0.05] | This slider helps in defining the minimum declaration- rate that can be used by all actors of artificial society |
| Safety Distance | [2 -10-; with increment of 1] | This slider helps in defining the safety distance between each actor. |
| Sonar Range | [1 -10-; with increment of 1] | This slider helps in defining the sonar range of each AV to find out the position and distance between neighboring Avs. |
| Metacognition | On/Off | This switch helps in defining that the simulation is in Random walk or social norms mode. |
| **Prospect Based Emotion i.e. Fear Generation Parameters** | Range | Description |
| Likelihood | [0 -1; with increment of 0.1] | This slider helps in defining the likelihood of accident perceived by AV |

| Desirability | [0 -1; with increment of 0.1] | This slider helps in defining the current desirability value of AV. |
| Ig | [0 -1; with increment of 0.1] | This slider helps in defining the current Ig value of AV. |

*5.2.2 Experimental design*

In this section, further experimental design has been proposed to perform the experiment 2 in proper manners.

### a) Experiments_TypeA

In this category of experiments, total five sets of experiments have been designed to test the non-social norms random walk based artificial society of AVs. In the first set of experiments, the Max Velocity Range parameter has been set to 0.8, which represents a high velocity of AVs. The Acceleration/ Deceleration Rate parameters are set to 0.1 along with the Safety Distance equal to 3. In the second set of experiments, the Max Velocity Range parameter has been set to 0.5, which represents a medium velocity of AVs. The Acceleration Rate parameter is set to 0.2 along with the Safety Distance equal to 2. In the third set of experiments, the Max velocity range is set to 0.3, which represent the low velocity of AVs. The Acceleration Rate and Declaration Rate parameters are both set to 0.1. In the fourth set of experiments, Acceleration and Deceleration Rates are set to 0.3 and the values of safety Distance and Sonar Range are set to 3. This set of experiment helps in measuring the performance of non-social norms random walk based artificial society of autonomous vehicles having equal safety distance and sonar range. In the fifth set of experiments, safety distance, and sonar range parameters are set to 1. This set of experiment helps in testing the behavior of AVs having equal low safety distance and sonar range. All of these sets of experiments have been executed using the behaviour space tool within the Netlogo 5.3 environment. Furthermore, each set of experiments has been repeated seven times and the total number of collisions along with their mean and standard deviation has been computed. The details of these 5 sets of experiments are presented in table 6, table 7, table 8, table 9, and table 10 respectively.

TABLE 6
EXPERIMENT TYPE_A SET 1: PARAMETERS AND THEIR VALUES

| Experiment No | Number of AVs | Min Velocity Range | Max Velocity Range | Acceleration Rate | Deceleration Rate | Safety Distance | Sonar Range |
|---|---|---|---|---|---|---|---|
| 1 | 10 | 0.14 | 0.8 | 0.1 | 0.1 | 3 | 2, 5 |
| 2 | 15 | 0.14 | 0.8 | 0.1 | 0.1 | 3 | 2, 5 |
| 3 | 20 | 0.14 | 0.8 | 0.1 | 0.1 | 3 | 2, 5 |
| 4 | 25 | 0.14 | 0.8 | 0.1 | 0.1 | 3 | 2, 5 |
| 5 | 30 | 0.14 | 0.8 | 0.1 | 0.1 | 3 | 2, 5 |

TABLE 7
EXPERIMENT TYPE_A SET 2: PARAMETERS AND THEIR VALUES

| Experiment No | Number of AVs | Min Velocity Range | Max Velocity Range | Acceleration Rate | Deceleration Rate | Safety Distance | Sonar Range |
|---|---|---|---|---|---|---|---|
| 1 | 10 | 0.14 | 0.5 | 0.2 | 0.2 | 2 | 2, 5 |
| 2 | 15 | 0.14 | 0.5 | 0.2 | 0.2 | 2 | 2, 5 |
| 3 | 20 | 0.14 | 0.5 | 0.2 | 0.2 | 2 | 2, 5 |
| 4 | 25 | 0.14 | 0.5 | 0.2 | 0.2 | 2 | 2, 5 |
| 5 | 30 | 0.14 | 0.5 | 0.2 | 0.2 | 2 | 2, 5 |

TABLE 8
EXPERIMENT TYPE_A SET 3: PARAMETERS AND THEIR VALUES

| Experiment No | Number of AVs | Min Velocity Range | Max Velocity Range | Acceleration Rate | Deceleration Rate | Safety Distance | Sonar Range |
|---|---|---|---|---|---|---|---|
| 1 | 10 | 0.14 | 0.3 | 0.1 | 0.1 | 2 | 2 |
| 2 | 15 | 0.14 | 0.3 | 0.1 | 0.1 | 2 | 2 |
| 3 | 20 | 0.14 | 0.3 | 0.1 | 0.1 | 2 | 2 |
| 4 | 25 | 0.14 | 0.3 | 0.1 | 0.1 | 2 | 2 |
| 5 | 30 | 0.14 | 0.3 | 0.1 | 0.1 | 2 | 2 |

TABLE 9
EXPERIMENT TYPE_A SET 4: PARAMETERS AND THEIR VALUES

| Experiment No | Number of AVs | Min Velocity Range | Max Velocity Range | Acceleration Rate | Deceleration Rate | Safety Distance | Sonar Range |
|---|---|---|---|---|---|---|---|
| 1 | 10 | 0.14 | 0.3 | 0.3 | 0.3 | 3 | 3 |
| 2 | 15 | 0.14 | 0.3 | 0.3 | 0.3 | 3 | 3 |
| 3 | 20 | 0.14 | 0.3 | 0.3 | 0.3 | 3 | 3 |
| 4 | 25 | 0.14 | 0.3 | 0.3 | 0.3 | 3 | 3 |
| 5 | 30 | 0.14 | 0.3 | 0.3 | 0.3 | 3 | 3 |

TABLE 10
EXPERIMENT TYPE_A SET 5: PARAMETERS AND THEIR VALUES

| Experiment No | Number of AVs | Min Velocity Range | Max Velocity Range | Acceleration Rate | Deceleration Rate | Safety Distance | Sonar Range |
|---|---|---|---|---|---|---|---|
| 1 | 10 | 0.14 | 0.3 | 0.1 | 0.1 | 1 | 1 |
| 2 | 15 | 0.14 | 0.3 | 0.1 | 0.1 | 1 | 1 |
| 3 | 20 | 0.14 | 0.3 | 0.1 | 0.1 | 1 | 1 |
| 4 | 25 | 0.14 | 0.3 | 0.1 | 0.1 | 1 | 1 |
| 5 | 30 | 0.14 | 0.3 | 0.1 | 0.1 | 1 | 1 |

  *b) Experiments_TypeB*

In this category of experiments, total five sets of experiments in parallel the Experiments_TypeA have been designed to test and compare the social norms and emotions inspired artificial society of AVs with non-social norms random walk based artificial society of AVs. These five sets of experiments are designed in parallel to the Type_A experiments. These sets of experiments have the same values of parameters as type_A experiments have. The additional parameter added in these experiments is the variables of fear, which help in computing the intensity of fear. Table 11 through 15 presents these 5 sets of experiments respectively.

TABLE 11
EXPERIMENT TYPE_B SET 1: PARAMETERS AND THEIR VALUES

| Experiment No | Number of AVs | Min Velocity Range | Max Velocity Range | Acceleration Rate | Deceleration Rate | Safety Distance | Sonar Range | LI, UD, Ig |
|---|---|---|---|---|---|---|---|---|
| 1 | 10 | 0.14 | 0.8 | 0.1 | 0.1 | 3 | 2 , 5 | 0.1-0.1-1 |
| 2 | 15 | 0.14 | 0.8 | 0.1 | 0.1 | 3 | 2 , 5 | 0.1-0.1-1 |
| 3 | 20 | 0.14 | 0.8 | 0.1 | 0.1 | 3 | 2 , 5 | 0.1-0.1-1 |
| 4 | 25 | 0.14 | 0.8 | 0.1 | 0.1 | 3 | 2, 5 | 0.1-0.1-1 |
| 5 | 30 | 0.14 | 0.8 | 0.1 | 0.1 | 3 | 2 , 5 | 0.1-0.1-1 |

TABLE 12
EXPERIMENT TYPE_B SET 2: PARAMETERS AND THEIR VALUES

| Experiment No | Number of AVs | Min Velocity Range | Max Velocity Range | Acceleration Rate | Deceleration Rate | Safety Distance | Sonar Range | LI, UD, Ig |
|---|---|---|---|---|---|---|---|---|
| 1 | 10 | 0.14 | 0.5 | 0.2 | 0.2 | 2 | 2, 5 | 0.1-0.1-1 |
| 2 | 15 | 0.14 | 0.5 | 0.2 | 0.2 | 2 | 2, 5 | 0.1-0.1-1 |
| 3 | 20 | 0.14 | 0.5 | 0.2 | 0.2 | 2 | 2, 5 | 0.1-0.1-1 |
| 4 | 25 | 0.14 | 0.5 | 0.2 | 0.2 | 2 | 2, 5 | 0.1-0.1-1 |
| 5 | 30 | 0.14 | 0.5 | 0.2 | 0.2 | 2 | 2, 5 | 0.1-0.1-1 |

TABLE 13
EXPERIMENT TYPE_B SET 3: PARAMETERS AND THEIR VALUES

| Experiment No | Number of AVs | Min Velocity Range | Max Velocity Range | Acceleration Rate | Deceleration Rate | Safety Distance | Sonar Range | LI, UD, Ig |
|---|---|---|---|---|---|---|---|---|
| 1 | 10 | 0.14 | 0.3 | 0.1 | 0.1 | 2 | 2 | 0.1-0.1-1 |
| 2 | 15 | 0.14 | 0.3 | 0.1 | 0.1 | 2 | 2 | 0.1-0.1-1 |
| 3 | 20 | 0.14 | 0.3 | 0.1 | 0.1 | 2 | 2 | 0.1-0.1-1 |
| 4 | 25 | 0.14 | 0.3 | 0.1 | 0.1 | 2 | 2 | 0.1-0.1-1 |
| 5 | 30 | 0.14 | 0.3 | 0.1 | 0.1 | 2 | 2 | 0.1-0.1-1 |

TABLE 14
EXPERIMENT TYPE_B SET 4: PARAMETERS AND THEIR VALUES

| Experiment No | Number of AVs | Min Velocity Range | Max Velocity Range | Acceleration Rate | Deceleration Rate | Safety Distance | Sonar Range | LI, UD, Ig |
|---|---|---|---|---|---|---|---|---|
| 1 | 10 | 0.14 | 0.3 | 0.3 | 0.3 | 3 | 3 | 0.1-0.1-1 |
| 2 | 15 | 0.14 | 0.3 | 0.3 | 0.3 | 3 | 3 | 0.1-0.1-1 |

| 3 | 20 | 0.14 | 0.3 | 0.3 | 0.3 | 3 | 3 | 0.1-0.1-1 |
| 4 | 25 | 0.14 | 0.3 | 0.3 | 0.3 | 3 | 3 | 0.1-0.1-1 |
| 5 | 30 | 0.14 | 0.3 | 0.3 | 0.3 | 3 | 3 | 0.1-0.1-1 |

TABLE 15
EXPERIMENT TYPE_B SET 5: PARAMETERS AND THEIR VALUES

| Experiment No | Number of AVs | Min Velocity Range | Max Velocity Range | Acceleration Rate | Deceleration Rate | Safety Distance | Sonar Range | LI, UD, Ig |
|---|---|---|---|---|---|---|---|---|
| 1 | 10 | 0.14 | 0.3 | 0.1 | 0.1 | 1 | 1 | 0.1-0.1-1 |
| 2 | 15 | 0.14 | 0.3 | 0.1 | 0.1 | 1 | 1 | 0.1-0.1-1 |
| 3 | 20 | 0.14 | 0.3 | 0.1 | 0.1 | 1 | 1 | 0.1-0.1-1 |
| 4 | 25 | 0.14 | 0.3 | 0.1 | 0.1 | 1 | 1 | 0.1-0.1-1 |
| 5 | 30 | 0.14 | 0.3 | 0.1 | 0.1 | 1 | 1 | 0.1-0.1-1 |

## 6-Results and Discussion

This section elaborates the detailed discussion according to the results achieved for experiment1 and experiment 2.

### 6.1 Experiment 1

Criado et al. [8], have utilized prospect based emotions defined by the OCC model to enforce the agents to obey the social norms. However, the authors have not proposed any proper mechanism, which helps to quantify the different intensities of fear. Furthermore, the authors have considered only Desirability and likelihood variables, whereas ignoring $I_g$ variable. In addition, the values of desirability and Likelihood variable are just supposed between [-1 1] without providing any justification. Our approach is better than [8] in this regard that we have used fuzzy logic to compute the numeric values of *Fear* variables (Likelihood, Desirability, $I_g$) to computer ***Fear Potential*** and then ***Fear Intensity*** has been computed using the proper algorithm defined by the inventors of the OCC model [25].

Table 1 shows the quantitative values of undesirability from very low (VL) to very high (VH). The terms VLD, LD, MD, HD, and VHD are the acronyms of very low desirability, low desirability, medium desirability, high desirability and very high desirability respectively. If the agent has a value between 0-0.24 for its undesirability of an event, then it can be interpreted as the very low undesirability. However, from an abstract analysis, it can be noted that due to the fuzzy nature of the emotion fear the boundary of one intensity level mixes in the boundary of another intensity level. Hence, the intensity levels lie between 0.24 and 0.5 will be interpreted as low undesirability and lower than these values as the very low undesirability. In the same way, the other intensity levels of undesirability variable can be interpreted.

In the same way, Table 2 and table 3 are showing the five quantitative values for finding the different intensity levels of likelihood and Ig variables.

These quantitative values of Desirability, Likelihood and Ig are presented in table 16, 17 and 18 respectively. These values are then provided to the EEC_Agent for computing different intensities of fear in the next section by following the proposed SimConnector design.

TABLE 16
Quantitative Values of Five Intensity levels of Desirable Variable

| VLD | LD | MD | HD | VHD |
| --- | --- | --- | --- | --- |
| 0-0.24 | 0.1-0.5 | 0.25-0.73 | 0.51-0.9 | 0.76-1 |

TABLE 17
QUANTITATIVE VALUES OF FIVE INTENSITY LEVELS OF LIKELIHOOD VARIABLE

| VLL | LL | ML | HL | VHL |
| --- | --- | --- | --- | --- |
| 0-0.24 | 0.1-0.5 | 0.25-0.73 | 0.51-0.9 | 0.76-1 |

TABLE 18
Quantitative Values of Five Intensity levels of Global Variable (Ig)

| VLIg | LIg | MIg | HIg | VIg |
| --- | --- | --- | --- | --- |
| 0-0.24 | 0.1-0.5 | 0.25-0.73 | 0.51-0.9 | 0.76-1 |

## 6.2 Experiment 2

*7.2.1 The Results: Experiment_TypeA set 1 Vs Experiment_TypeB set 1 and Experiment_TypeA set 2 Vs Experiment_TypeB set 2*

The results of both experiments_TypeA set 1 and experiments_Type B set1 are presented in the form of average accidents along with standard deviation in table 19. From the results, it can be seen that there is a high average of accidents in case of non-social norms random walk based artificial society of AVs as compared to the social norms and emotions based artificial society of AVs. For example, the average accidents performed by non-social norms random walk based are 48.63 for 10 AVs. Comparatively, 2.57 are the average accidents performed by social norms and emotions based technique. In the same way, for 30 AVs total average accidents by non-social norms random walk are 305. 43 and 59. 35 by the social norms and emotions based technique. From the results, another interesting phenomenon can be observed that the average accidents in both techniques are gradually increasing as the number of AVs is increasing. Figure 6 (A) is representing the graphical representation of the results of table 19.

In comparison to the table 19, table 20 has been presented. Table 20 presents the results of both experiment_TypeA set2 and experiment_TypeB set 2 in the form of average accidents along with standard deviation. Before discussing the results of table 20 it would be interesting to perform the comparison of table 19 and table 20. Experiments_TypeA set 1 and experiments_TypeB set1 have a high maximum velocity range, i.e. 0.8 with acceleration and deceleration rate 0.1. Whereas, experiments_TypeA set2 and experiments_TypeB set 2 have medium maximum velocity range, i.e. 0.5 with acceleration and deceleration rates 0.2. From table 20, it can be seen that average accidents have been decreased due to medium maximum velocity range as compared to the average accidents presented in table 19 having a high maximum velocity range. For example, in table 19 and 20 for social norms and emotions based technique, the average accidents performed by 10 AVs are 2.57 and 44.35 respectively. In the same way, in the case of 30 AVs, the average accidents performed by social norms and emotions based technique are 59.35 and 24.065 respectively.

From the table 20, it can be seen that social norms and emotions based technique have less number of collisions as compared to the non-social norms random walk based technique. For example, the average accidents performed by social norms and emotions based artificial society are 6.35 for 20 AVs. Whereas, for the same number of AVs the average accidents performed by non-social norms based artificial society are 147.03. Figure 6 (B) is representing the graphical representation of the results of table 20.

TABLE 19
Set 1 Type A & B Experiments Results

| No. of AVs | Social norms and Emotions Based | | Nonsocial norms Random walk based | |
|---|---|---|---|---|
| | Mean | Stdev | Mean | Stdev |
| 10 AVs | 2.573837 | 1.597861 | 48.63886 | 10.05186 |
| 15 AVs | 7.317401 | 3.12966 | 95.57575 | 13.87058 |
| 20 AVs | 16.17935 | 6.106475 | 152.584 | 18.13361 |
| 25 AVs | 32.20731 | 11.09385 | 222.3648 | 21.98538 |
| 30 AVs | 59.35412 | 17.14856 | 305.439 | 25.01929 |

TABLE 20
Set 2 Type A & B Experiments Results

| No. of AVs | Social norms and Emotions Based | | Non-social norms Random walk based | |
|---|---|---|---|---|
| | Mean | Stdev | Mean | Stdev |
| 10 AVs | 0.83308 | 0.75219454 | 44.35492 | 10.41238 |
| 15 AVs | 2.688164 | 1.50878434 | 92.14185 | 12.76792 |
| 20 AVs | 6.358672 | 2.40534634 | 147.0383 | 15.39402 |
| 25 AVs | 13.37643 | 3.34040996 | 207.6984 | 17.56855 |
| 30 AVs | 24.06544 | 4.45493639 | 276.0833 | 19.05506 |

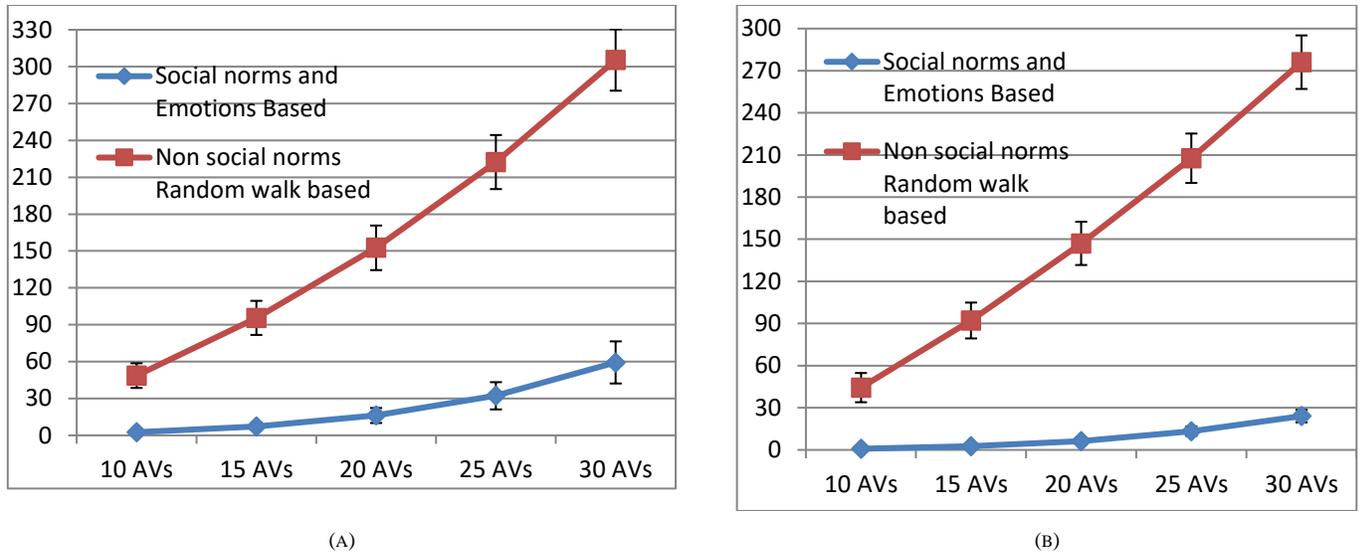

(A)                                (B)

Fig. 6 Graphical representation of the results (A) Experiment_TypeA set 1 Vs Experiment_TypeB set 1 (B) Experiment_TypeA set 2 Vs Experiment_TypeB set 2

In the same way, for the 10, 20, 25, and 30 number of AVs, average number of collisions by the social norms and emotions based artificial society of AVs are less than non-social norms and random walk based artificial society of AVs.

It would be interesting to present here the analysis of TypeB set 1 and set 2 experiments with Type B set 3 experiments. From the comparative study of table 19, 20 and 21 it can be seen that Type B set 3 results are better than Type B set1 and set 2 experiments. For example, for 30 AVs, the average collisions performed by set1 and set 2 are 59.35 and 24.06. Whereas there are only 14.69 collisions on average by TypeB set 3. Hence, it can be concluded that social norms and emotions based artificial society of AVs can have less number of collisions by adapting low maximum velocity range i.e. 0.3 and both safety and sonar distances equal to 2. Figure 7 (C) and (D) are the graphical representations of table 21 and 22 respectively.

TABLE 21
Set 3 Type A & B Experiments Results

| No. of AVs | Social norms and Emotions Based | | Nonsocial norms Random walk based | |
|---|---|---|---|---|
| | Mean | Stdev | Mean | Stdev |
| 10 AVs | 0.785513 | 0.916095 | 22.66486 | 10.09421 |
| 15 AVs | 2.268417 | 1.664854 | 55.82655 | 13.73423 |
| 20 AVs | 4.767293 | 2.38503 | 101.9607 | 17.59697 |
| 25 AVs | 8.728624 | 3.078323 | 158.7121 | 20.06934 |
| 30 AVs | 14.69935 | 3.936857 | 222.0574 | 22.18164 |

TABLE 22
Set 4 Type A & B Experiments Results

| No. of AVs | Social norms and Emotions Based | | Non-social norms Random walk based | |
|---|---|---|---|---|
| | Mean | Stdev | Mean | Stdev |
| 10 AVs | 1.889218 | 2.067026 | 25.12701 | 9.707391 |
| 15 AVs | 6.042479 | 3.723965 | 60.30161 | 13.30325 |
| 20 AVs | 14.54727 | 5.010924 | 106.1374 | 14.68821 |
| 25 AVs | 28.01678 | 6.281523 | 158.8475 | 17.78358 |
| 30 AVs | 44.53196 | 7.017011 | 219.4805 | 18.46683 |

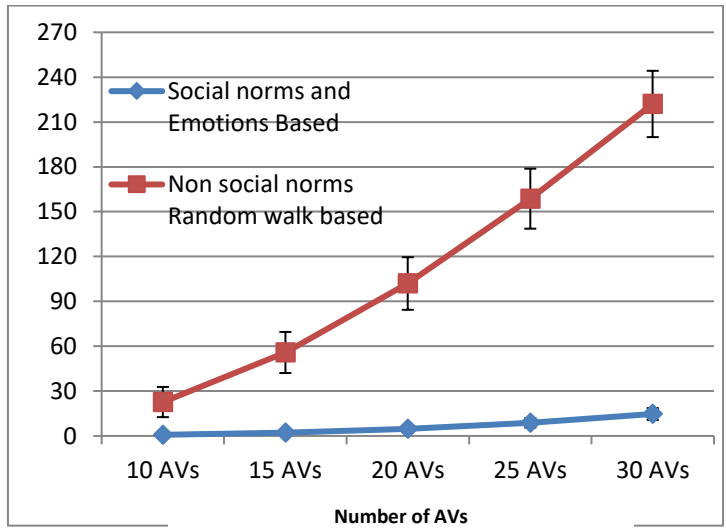

(C)

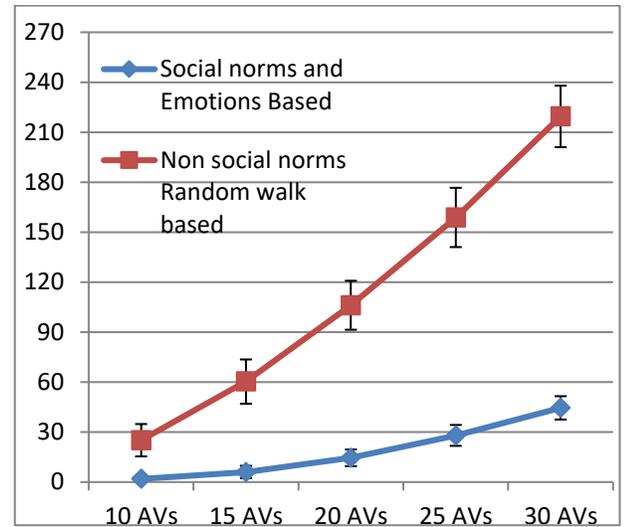

(D)

Fig. 7 Graphical representation of the results (C)_TypeA set 3 Vs Experiment_TypeB set 3 (D) Experiment_TypeA set 4 Vs Experiment_TypeB set 4

*6.2.3 The Results: Experiment_TypeA and TypeB set 3 Vs Experiment_TypeA set 5 Vs Experiment_TypeA and Type_B set 5*

Both sets 3 and 5 of Experiments_Type A and B have the same values of parameters expect safety distance and sonar range. In the set 3, both safety and sonar range are set to 3. Whereas for the set 5, both of these parameters are set to 1. From the comparative analysis, it can be seen that the set 3 has less number of collisions for 10, 15, 20 and 25 AVs. However, for 30 AVs, set 5 haS less number of accidents. From the trend line shown in figure 8 (E) for experiments_TypeA set 3, it can be seen that the number of collisions is increasing gradually as the number of AVs is increasing. In contrast to the trend line shown in figure 8 (F) for experiments_TypeA set 3 presents a gradual increase for 10, 15, 20, and 25 AVs but then suddenly drop down for 30 AVs. Hence, it means that the set 3 provides optimal operational parameters for the artificial society of AVs within the range 1 to 25. Whereas, for 30 AVs experiment no 5 of set 5 is the optimal option. From these results, we can also deduce that for higher AVs, small and equal safety distance and sonar range parameters are most optimal one.

If we see the results of table 23 then it is obvious that social norms and emotions based artificial society of AVs have less number of collisions for all numbers of AVs as compared to the non-social norms and emotions based artificial society of AVs.

*6.2.4 Analysis of most optimal Sonar Range Vs. Safety Distance for less number of collisions in Social norms and Emotions Based artificial society of AVs*

Table 24 presents the number of collisions for different number of AVs according to different safety distances and sonar ranges in experiments_TypeA set1 to set5. From the results, it can be seen that when safety distance and sonar range parameters having values (1, 1) and (2, 2) respectively, the average number of collisions is lesser. For example, for 10 AVs with safety distance and sonar range parameters set to (1, 1) and (1, 2) the average number of collisions are 0.79 and 0.78 respectively as compared to the 3.01, 1.88 and 2.12 for sonar range and safety distance parameters set to (3, 2), (3, 3), and (3, 5) respectively. From the set4, experiments_TypeA with safety distance and sonar range parameters set to (3, 2) the average number of collisions is higher than all other experiments. Its reason is smaller safety distance than the sonar range. It means that when the AV has higher safety

TABLE 21
Set 3 Type A & B Experiments Results

| No. of AVs | Social norms and Emotions Based | | Nonsocial norms Random walk based | |
|---|---|---|---|---|
| | Mean | Stdev | Mean | Stdev |
| 10 AVs | 0.785513 | 0.916095 | 22.66486 | 10.09421 |
| 15 AVs | 2.268417 | 1.664854 | 55.82655 | 13.73423 |
| 20 AVs | 4.767293 | 2.38503 | 101.9607 | 17.59697 |
| 25 AVs | 8.728624 | 3.078323 | 158.7121 | 20.06934 |
| 30 AVs | 14.69935 | 3.936857 | 222.0574 | 22.18164 |

TABLE 23
Set 5 Type A & B Experiments Results

| No. of AVs | Social norms and Emotions Based | | Nonsocial norms Random walk based | |
|---|---|---|---|---|
| | Mean | Stdev | Mean | Stdev |
| 10 AVs | 0.791342 | 0.576955 | 37.75169 | 11.27174 |
| 15 AVs | 2.084906 | 0.999392 | 85.15609 | 14.30291 |
| 20 AVs | 7.152462 | 1.965884 | 145.9167 | 16.50659 |
| 25 AVs | 11.63587 | 2.745418 | 215.4169 | 18.68783 |
| 30 AVs | 4.069908 | 1.37926 | 294.7733 | 20.14137 |

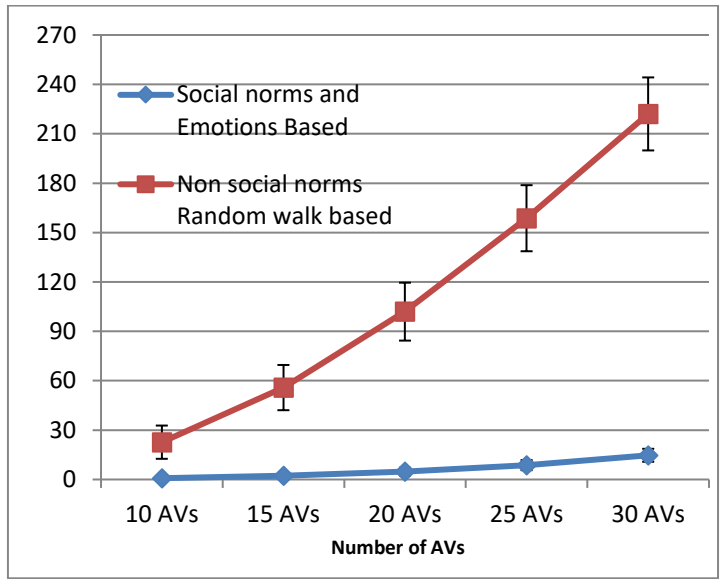

(E)

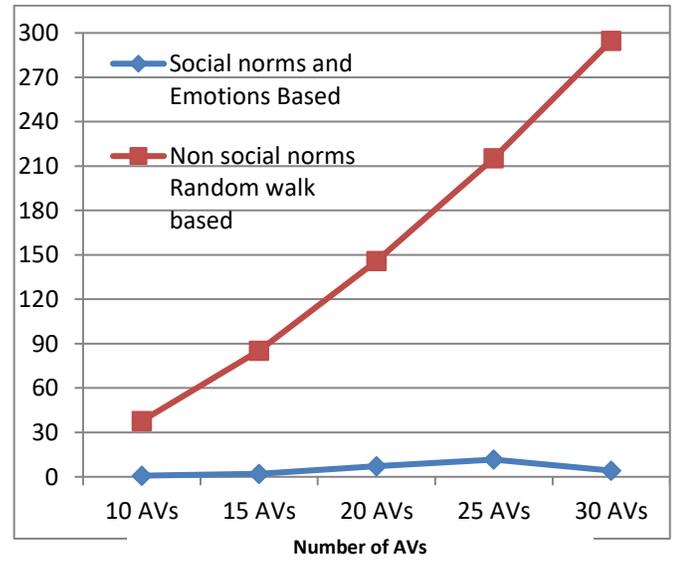

(F)

Fig. 8 Graphical representation of the results (E) Experiment_TypeA set 3 Vs Experiment_TypeB set 3 (F) Experiment_TypeA set 5 Vs Experiment_TypeB set 5

distance and low capability of detecting its neighbors then the number of collisions increases. Figure 9 presents the graphical representation of the results of table 24.

TABLE 24
Experimental results of TypeA set1- set 5 regarding different sonar ranges

| Safety range, Sonar distance | 10 AVs1 | 15 AVs | 20 AVs | 25 AVs | 30 AVs |
|---|---|---|---|---|---|
| 1, 1 | 0.791342 | 2.084906 | 7.152462 | 11.63587 | 4.069908 |
| 2, 2 | 0.785513 | 2.268417 | 4.767293 | 8.728624 | 14.69935 |
| **2, 5** | 0.814496 | 2.711523 | 6.390049 | 13.36398 | 24.04687 |
| **3, 2** | 3.01972 | 8.936398 | 20.14434 | 40.31569 | 72.42839 |
| **3, 3** | 1.889218 | 6.042479 | 14.54727 | 28.01678 | 44.53196 |
| **3, 5** | 2.127955 | 5.698404 | 12.21435 | 24.09892 | 46.27984 |

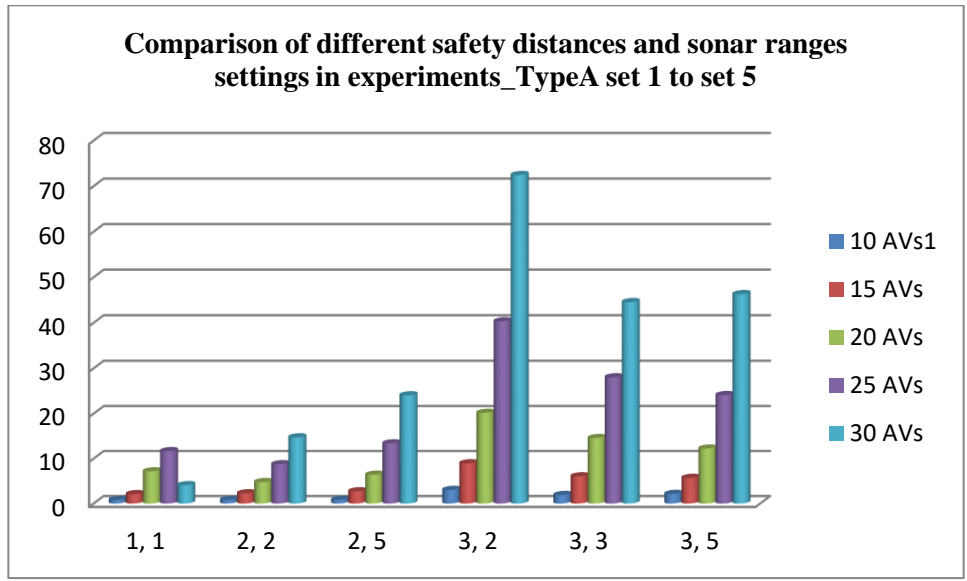

FIG. 9 Graphical representation of Results regarding different sonar ranges given in Experiments: TypeA set1 - set 5

## 7-Conclusion

The paper has been written in the context of proposing a novel collision avoidance solution for the AVs, when they will be the main players of the road traffic. In the near future, it has been assumed that AVs will be very common and people will delegate their driving powers to them. To answer the question that how AVs will be able to fulfill the expectations of humans in terms of safer road operations with less number of collisions and harmless interactions with each other , especially when human drivers have no role in their operations, this research work has been done. The answer has been provided through the human social life protocol, which lies in their core, humans, to interact with each other, avoiding the conflicts, and keeping the social society in equilibrium. The key is following social norms under the influence of primary emotions. Furthermore, the simulation results have provided optimal parameters, like optimal sonar range and different optimal speeds suitable for avoiding the road collisions in different road traffic situations. This research work might be suitable for AV vendors to reinvent the autopilot design, in terms of including social norms, emotions and optimal operating parameters. Hopefully it will make AVs capable to cope with the current dilemma that how the AVs make themselves more trustworthy in terms of safe travelling.

**References**

[1] F. Riaz and M. A. Niazi, "Road collisions avoidance using vehicular cyber-physical systems: a taxonomy and review," *Complex Adaptive Systems Modeling,* vol. 4, p. 1, 2016.
[2] T. Litman, "Autonomous Vehicle Implementation Predictions," *Victoria Transport Policy Institute,* vol. 28, 2014.
[3] A. C. Mersky and C. Samaras, "Fuel economy testing of autonomous vehicles," *Transportation Research Part C: Emerging Technologies,* vol. 65, pp. 31-48, 2016.
[4] B. F. Malle, "Integrating robot ethics and machine morality: the study and design of moral competence in robots," *Ethics and Information Technology,* pp. 1-14, 2015.
[5] A. Rakotonirainy, R. Schroeter, and A. Soro, "Three social car visions to improve driver behaviour," *Pervasive and Mobile Computing,* vol. 14, pp. 147-160, 2014.
[6] W. Kumfer and R. Burgess, "Investigation into the Role of Rational Ethics in Crashes of Automated Vehicles," *Transportation Research Record: Journal of the Transportation Research Board,* pp. 130-136, 2015.
[7] J. Elster, "Social norms and economic theory," in *Culture and Politics*, ed: Springer, 2000, pp. 363-380.
[8] N. Criado, E. Argente, P. Noriega, and V. Botti, "Human-inspired model for norm compliance decision making," *Information Sciences,* vol. 245, pp. 218-239, 2013.
[9] A. Staller and P. Petta, "Introducing emotions into the computational study of social norms: A first evaluation," *Journal of artificial societies and social simulation,* vol. 4, pp. U27-U60, 2001.
[10] J. C. Gerdes and S. M. Thornton, "Implementable Ethics for Autonomous Vehicles," in *Autonomous Driving*, ed: Springer, 2016, pp. 87-102.
[11] H. T. Tavani, *Ethics and technology: Controversies, questions, and strategies for ethical computing*: John Wiley & Sons, 2011.
[12] M. Van de Voort, W. Pieters, and L. Consoli, "Refining the ethics of computer-made decisions: a classification of moral mediation by ubiquitous machines," *Ethics and Information Technology,* vol. 17, pp. 41-56, 2015.
[13] A. Etzioni and O. Etzioni, "AI assisted ethics," *Ethics and Information Technology,* vol. 18, pp. 149-156, 2016.
[14] T. Schmidt, R. Philipsen, and M. Ziefle, "User Diverse Privacy Requirements for V2X-Technology," 2016.
[15] J. Gogoll and J. F. Müller, "Autonomous cars: in favor of a mandatory ethics setting," *Science and Engineering Ethics,* pp. 1-20, 2016.
[16] A. R. Hochschild, "Emotion work, feeling rules, and social structure," *American journal of sociology,* pp. 551-575, 1979.
[17] A. Ortony, G. L. Clore, and A. Collins, *The cognitive structure of emotions*: Cambridge university press, 1990.
[18] W. G. Najm, J. D. Smith, and M. Yanagisawa, "Pre-crash scenario typology for crash avoidance research," in *DOT HS*, 2007.
[19] M. S. El-Nasr and J. Yen, "Agents, emotional intelligence and fuzzy logic," in *Fuzzy Information Processing Society-NAFIPS, 1998 Conference of the North American*, 1998, pp. 301-305.
[20] M. S. El-Nasr, J. Yen, and T. R. Ioerger, "Flame—fuzzy logic adaptive model of emotions," *Autonomous Agents and Multi-agent systems,* vol. 3, pp. 219-257, 2000.
[21] A. Aly and A. Tapus, "An Online Fuzzy-Based Approach for Human Emotions Detection: An Overview on the Human Cognitive Model of Understanding and Generating Multimodal Actions," in *Intelligent Assistive Robots*, ed: Springer, 2015, pp. 185-212.
[22] R. L. Mandryk and M. S. Atkins, "A fuzzy physiological approach for continuously modeling emotion during interaction with play technologies," *International journal of human-computer studies,* vol. 65, pp. 329-347, 2007.
[23] C. E. Izard, *Human emotions*: Springer Science & Business Media, 2013.
[24] M. A. Niazi, "Emergence of a Snake-Like Structure in Mobile Distributed Agents: An Exploratory Agent-Based Modeling Approach," *The Scientific World Journal,* vol. 2014, 2014.

**Appendix A**

*The Intensity of global variables ($I_g$):*

To compute the intensity of $I_g$ variable, five linguistic tokens VLIG, LIG, MIG, HIG and VHIG were defined, which represent the very low intensity of Low likelihood, medium likelihood, high likelihood and the very high likelihood respectively. The linguistic tokens of $I_g$ are presented in table 25.

TABLE 25
$I_G$ - LINGUISTIC TOKENS AND THEIR DESCRIPTION

| Linguistic Tokens | Description |
|---|---|
| VLIG | Very low intensity of global variable |
| LIG | Low intensity of global variable |
| MIG | Medium intensity of global variable |
| HIG | High intensity of global variable |
| VHIG | Very high intensity of global variable |

The intensity of global variable depends on proximity and a sense of reality variables. Twenty-five rules were defined to obtain the value of the variable likelihood; these rules are given in table 26.

TABLE 26
IG - LIKELIHOOD FUZZY INFERENCE RULES

| If Sense of Reality is | And Proximity is | Then Intensity of Goal will be |
|---|---|---|
| VLSOR | About to | MIG |
| VLSOR | Going to | MIG |
| VLSOR | MChance | LIG |
| VLSOR | LChance | VLIG |
| VLSOR | NChance | VLIG |
| LSOR | About to | HIG |
| LSOR | Going to | MIG |
| LSOR | MChance | MIG |
| LSOR | LChance | LIG |
| LSOR | NChance | VLIG |
| MSOR | About to | HIG |
| MSOR | Going to | HIG |
| MSOR | MChance | MIG |
| MSOR | LChance | LIG |
| MSOR | NChance | VLIG |
| HSOR | About to | VHIG |
| HSOR | Going to | HIG |
| HSOR | MChance | MIG |
| HSOR | LChance | LIG |
| HSOR | NChance | VLIG |
| VHSOR | About to | VHIG |
| VHSOR | Going to | VHIG |

| VHSOR | MChance | HIG |
|-------|---------|-----|
| VHSOR | LChance | HIG |
| VHSOR | NChance | MIG |

*Experiment A. Computing Undesirability*

According to the OCC model, desirability is a local variable, which affects only event, and agent-based emotions. The desirability variable further comprises two sub-variables: First, one is the importance of the goal and the second one is the achievement of the goal.

The effects of these two sub-variables in computing the desirability can be seen in the following scenario. Suppose that the goal of AV is reaching its destination on time. Suddenly the battery of the AV gets down. Now here the undesirability of the event can have more than one values. We are just representing here two cases. If the importance of goal is very high and it has traveled only 30 % of the distance towards its destination, then undesirability of the said event will be very high. In the second case, if the importance of goal is very low and it has achieved 100 % of an assigned task (battery gets down after reaching its destination) then the undesirability of the event will be very low. The main simulation screen of computing desirability (undesirability in the case of fear) is shown in figure 10. The screen is showing two input variables and one output variable. The input variables are the importance of Goal (ImpGoal), achievement of the goal (AchGoal) and the output variable is Undesirability.

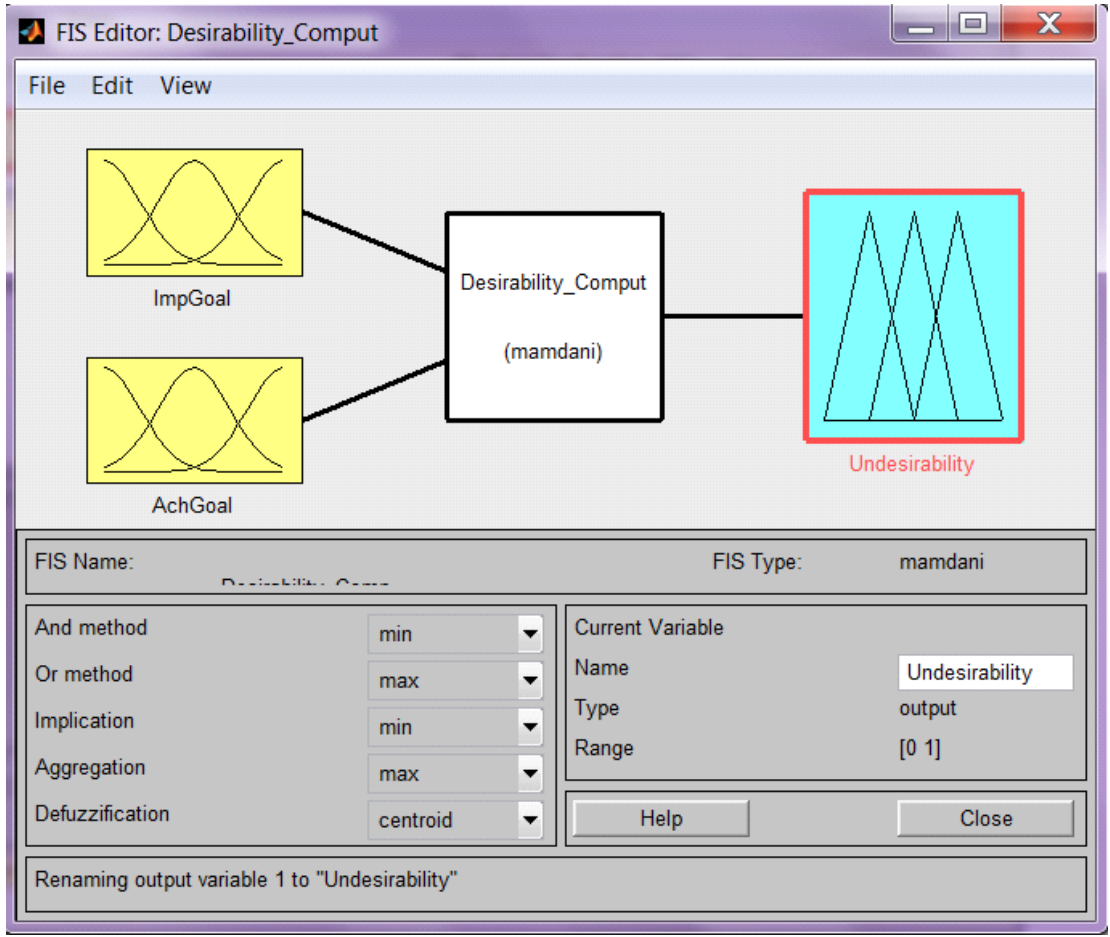

Fig. 10 Main simulation screen for Desirability computation

To compute undesirability trigonometric function (trimf) has been used with linguistic tokens VLUD, LUD, MUD, HUD and VHUD which represent Very low Undesirable, Low Undesirable, Medium Undesirable, High Undesirable and Very High Undesirable respectively to represent different intensity levels of undesirability. Twenty-five rules were defined to obtain the value of the variable undesirability; these rules are given in table 27.

*Validation of fuzzy logic rules for computing the Undesirability*

The validation of undesirability fuzzy rules has been performed in rule view of FIS editor. Rule viewer was provided random values for different linguistic tokens and in the result, fuzzy inference system computed different intensities of undesirability. To cross check the outcomes hand trace mechanism has been adopted, which further validated the outcomes of different undesirability values shown in table 28. In test 1, it can be seen that the input variables ImpGoal, AchGoal have values 0.1 and 0.5, which lies in the very low range and medium range respectively. In result, the FIS system computes low undesirability i.e. 0.25, which is correct. In the same way in test 7 linguistic tokens medium importance of goal i.e. MImpG, medium achieved goal MAG has values 0.56 and 0.5, which lies in the medium range. In a result, the FIS system computes medium intensity of undesirability i.e.

TABLE 27
DESIRABILITY- LIKELIHOOD FUZZY INFERENCE RULES

| If Importance of Goal is | And Achievement of goal is | Then undesirability will be |
|---|---|---|
| VLImpG | NAG | MUD |
| VLImpG | LAG | LUD |
| VLImpG | MAG | LUD |
| VLImpG | HAG | VLUD |
| VLImpG | HFAG | VLUD |
| LImpG | NAG | MUD |
| LImpG | LAG | MUD |
| LImpG | MAG | LUD |
| LImpG | HAG | VLUD |
| LImpG | VHFAG | VLUD |
| MImpG | NAG | HUD |
| MImpG | LAG | MUD |
| MImpG | MAG | MUD |
| MImpG | HAG | LUD |
| MImpG | VHFAG | LUD |
| HImpG | NAG | VHUD |
| HImpG | LAG | HUD |
| HImpG | MAG | HUD |
| HImpG | HAG | MUD |
| HImpG | VHFAG | VHUD |
| VHImpG | NAG | VHUD |
| VHImpG | LAG | HUD |
| VHImpG | MAG | HUD |
| VHImpG | HAG | HUD |
| *VHImpG* | *VHFAG* | *MUD* |

0.567, which is correct. In the same way, other validation results can be cross-checked using hand tracing mechanism.

The following table shows that different values for ImpGoal and AchGoal were entered as input and each time output value of the Undesirability variable is according to the rules.

TABLE 28
VALIDATION OF FUZZY LOGIC RULES FOR COMPUTING THE UNDESIRABILITY

| No. Of Tests | ImpGoal | AchGoal | Undesirability |
|---|---|---|---|
| 1 | 0.1(VLImpG) | 0.5(MAG) | 0.25(LUD) |
| 2 | 0.2(VLImpG) | 1.0(VHAG) | 0.08(VLUD) |
| 3 | 0.27(LImpG) | 0(NAG) | 0.52(MUD) |
| 4 | O.30(LImpG) | 0.5(MAG) | 0.31(LUD) |
| 5 | 0.4(LImpG) | 1.0(VHAG) | 0.09(VLUD) |
| 6 | 0.5(MImpG) | 0(NAG) | 0.74(HUD) |
| 7 | 0.56(MImpG) | 0.5(MAG) | 0.567(MUD) |
| 8 | 0.6(MImpG) | 1.0(VHAG) | 0.09(VLUD) |
| 9 | 0.8(HImpG) | 0(NAG) | 0.91(VHUD) |
| 10 | 0.85(HImpG) | 0.5(MAG) | 0.746(HUD) |
| 11 | 0.79(HImpG) | 1.0(VHAG) | 0.085(VLUD) |
| 12 | 0.96(VHImpG) | 0(NAG) | 0.917(VHUD) |
| 13 | 0.98(VHImpG) | 0.5(MAG) | 0.747(HUD) |
| 14 | 1.0(VHImpG) | 1.0(VHAG) | 0.08(VLUD) |

*Experiment B. Computing Likelihood*

The Likelihood of the event depends on the Distance and speed of the following and leading AVs. In our case, the Likelihood is representing TTA (Time To Avoid). For example, if the distance between two vehicles is low and their speed is in high range then it leads to the higher Likelihood of collision between these two vehicles. Therefore, the two variables, which affect the likelihood of an event, are; the first one is the distance between both AVs and the second one is the speed of Bullet AV.

The figure 11 is representing the main simulation screen utilized to compute the Likelihood variable. The screen is showing two input variables and one output variable. The input variables are Speed and Distance and the output variable is Likelihood as discussed above.

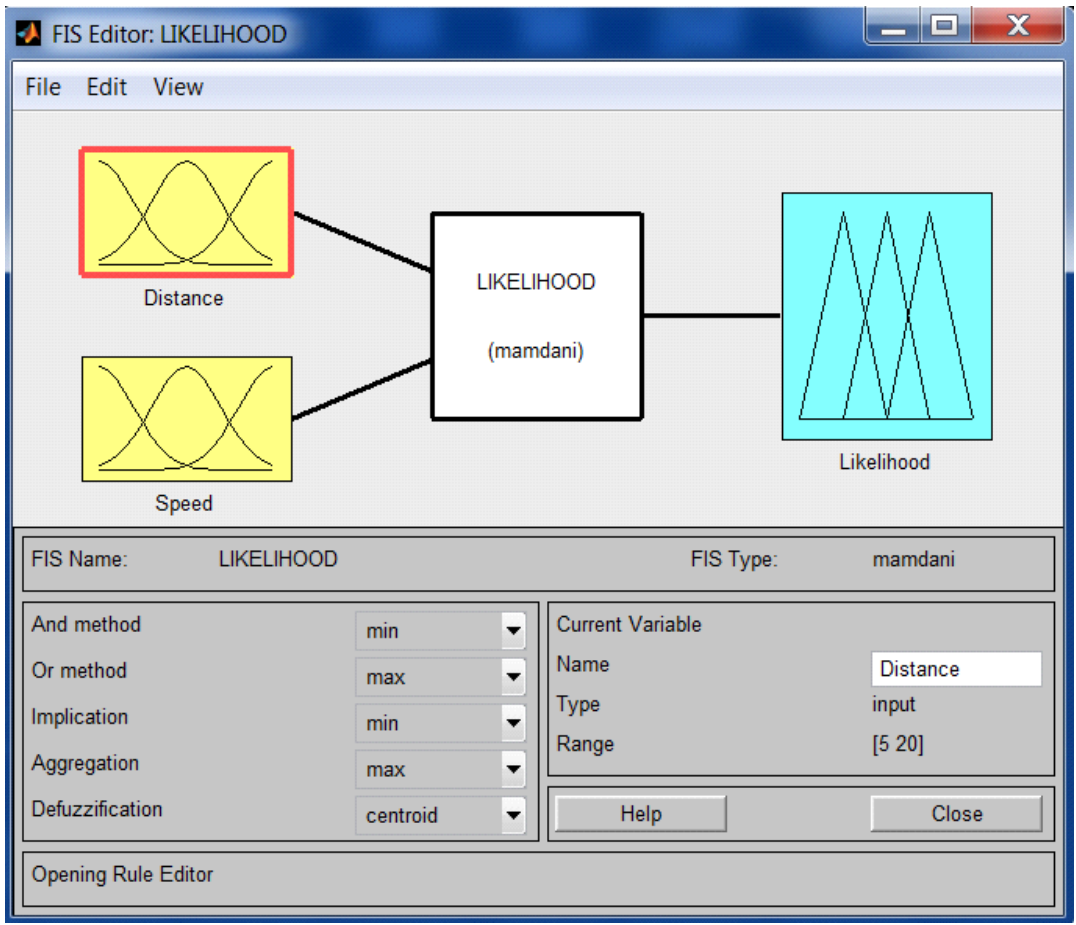

Fig. 11 Main simulation screen for Likelihood computation

*Experiment C. Intensity of Global Variable*

The intensity of global variable further depends on proximity and the sense of reality variables. The sense of reality is the scene interpretation by the sensing module of AV or the reality of the event on which AV believes or not. This variable has the global influence on the intensity of emotions. Proximity is the distance between the AVs. The proximity influences the intensity of emotions that can involve future situations. We have taken proximity here in spatial terms.

Figure 12 depicts the main simulation screen regarding the quantitative computation of Ig variable. Here the sense of reality and proximity are acting as two input variables to compute Ig.

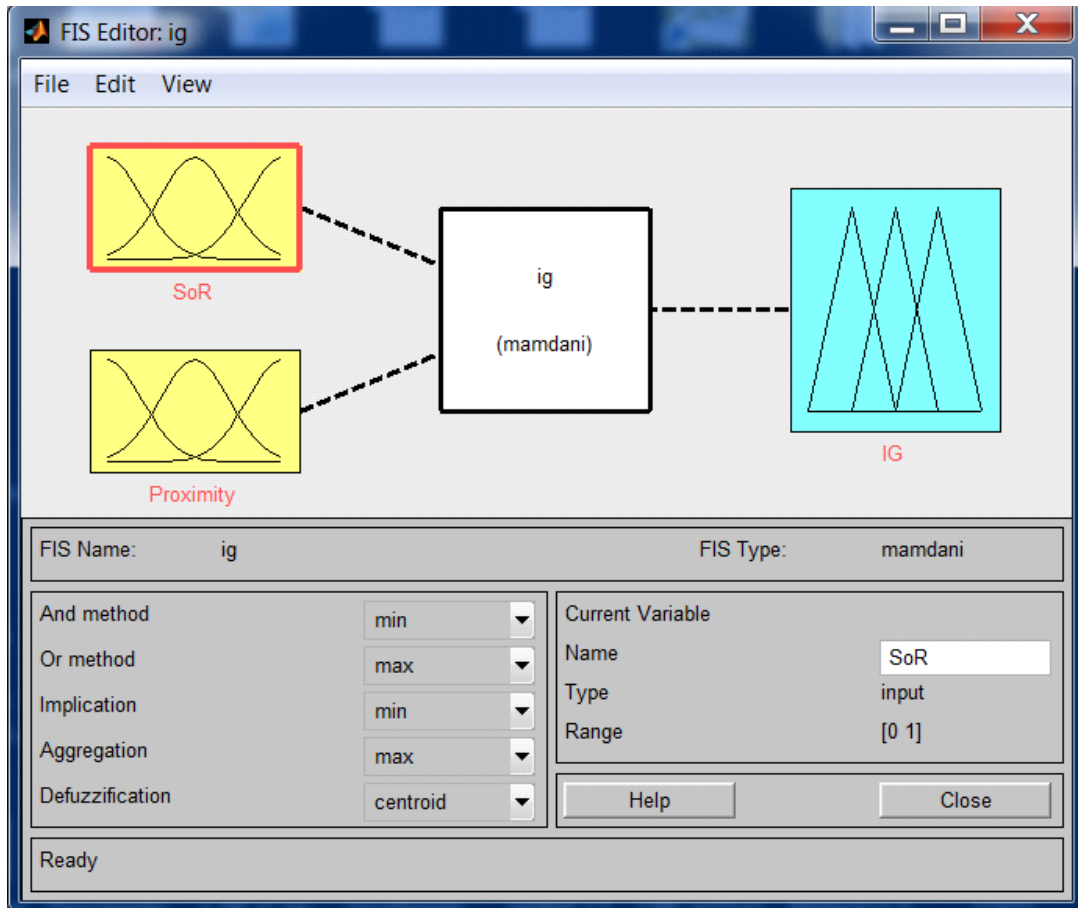

Fig.12 Main Simulation Screen of the Intensity of global variable